# The Function Transformation Omics – Funomics

Yongshuai Jiang[1,*] Jing Xu[1], Simeng Hu[1], Di Liu[1], Linna Zhao[1] , Xu Zhou[1]

1.College of Bioinformatics Science and Technology, Harbin Medical University, Harbin, 150086, China;
*To whom correspondence should be addressed: jiangyongshuai@gmail.com.

## Abstract

There are no two identical leaves in the world, so how to find effective markers or features to distinguish them is an important issue. Function transformation, such as f(x,y) and f(x,y,z), can transform two, three, or multiple input/observation variables (in biology, it generally refers to the observed/measured value of biomarkers, biological characteristics, or other indicators) into a new output variable (new characteristics or indicators). This provided us a chance to re-cognize objective things or relationships beyond the original measurements. For example, Body Mass Index, which transform weight and high into a new indicator $BMI=x/y^2$ (where x is weight and y is high), is commonly used in to gauge obesity. Here, we proposed a new system, Funomics (Function Transformation Omics), for understanding the world in a different perspective. Funome can be understood as a set of math functions consist of basic elementary functions (such as power functions and exponential functions) and basic mathematical operations (such as addition, subtraction). By scanning the whole Funome, researchers can identify some special functions (called handsome functions) which can generate the novel important output variable (characteristics or indicators). We also start "the Funome project" to develop novel methods, function library and analysis software for Funome studies. The Funome project will accelerate the discovery of new useful indicators or characteristics, will improve the utilization efficiency of directly measured data, and will enhance our ability to understand the world. The analysis tools and data resources about the Funome project can be found gradually at http://www.funome.com.

**Introduction**

An important goal in scientific research is to find some effective characteristics to distinguish objective things. Especially in the field of life and biomedical science, the researchers hope to find some biomarkers to detect disease, such as cancer[1,2], diabetes[3] and alzheimer's disease[4], or to test whether a treatment is working[5]. With the development of biotechnology, such as gene chip and sequencing technology[6,7], biomarkers of various omics (genome, transcriptome, proteome, epigenome, etc.)[8] have been widely and directly measured. However, the scanning of single markers can only provide limited understanding of complex disease.

Math function transformation is an effective way to transform two, three, or multiple observed/measured values of biomarkers (or biological characteristics) into a new useful characteristic/indicator. One example is Body Mass Index (BMI) which is an important indicator of body fat, and widely be used in scientific research[9,10]. Let $x$ is person's weight and $y$ is person's high, $BMI = f(x,y) = x/y^2$ is actually a function transformation which transform the person's weight and height to a new indicator. Another example is plasma amyloid-β (Aβ) biomarkers for Alzheimer's disease[11]. Akinori Nakamura et al. measured plasma amyloid-β biomarkers by immunoprecipitation coupled with mass spectrometry, and found Aβ$_{1-40}$/Aβ$_{1-42}$ ratio has a high performance to predict individual brain amyloid-β -positive or -negative status. Let $x$ is Aβ$_{1-40}$ and $y$ is Aβ$_{1-42}$, $F = f(x,y) = x/y$ is a function transformation which transform the Aβ$_{1-40}$ and Aβ$_{1-42}$ to a new indicator. The new indicator is a potentially clinically useful candidate for plasma biomarker as surrogates for brain Aβ burden.

There are a lot of classic examples, such as waist-hip ratio (WHR)[12], we will not enumerate out them one by one. In order to extend the application of function transformation, we propose a strategy, Funomics (Function Transformation Omics), to discover new useful indicators or characteristics by scanning a set of functions.

**The Funome Project**

*Funome*

A funome is a set of function transformations. In all kinds of function transformations, the math functions generated from the basic elementary function are the most easily understood by researchers. In the Funome Project, we mainly consider the common mathematical functions, such as

Two-Variable Functions (2v Funome), Three-Variable Functions (3v Funome), or some special function sets (special Funomes).

*The generation of functions*

All the functions in current version of 'the Funome Project' were generated from the basic elementary function, such as exponential function, power function, exponential function, logarithmic function, trigonometric function, anti-trigonometric function. Basic mathematical operations, such as addition, subtraction, multiplication, division, powers and compound, were used to procedure the math functions.

*Identifying the novel indicators/characteristics by scanning the funome*

An important task of 'the Funome Project' is to identify novel indicators/characteristics from the original observed/measured variables (biomarkers or biological characteristics). For some omics data, such as genome, transcriptome and proteome, we will scan the funome to find new indicators which can distinguish disease phenotypes or drug targets. The new indicators may constituted by several biomarkers/characteristics that are not significant.

*Handsome function*

In the Funome Project, the function ( $F = f(x, y, z,...)$ ) identified by scanning funome is called handsome function. The dependent variable, $F$, in handsome function can also be called handsome indicator/characteristic. The independent variable, $x, y, z,...$, of handsome function can be called handsome original variables. For example, $BMI = F = f(x, y) = x/y^2$, where $F = x/y^2$ is a handsome function, $F$ is a handsome indicator named BMI, $x, y$ (weight and height) are handsome original variables.

*Develop the Funome tools*

Another task of 'the Funome Project' is to develop some tools to scan the funome and identify handsome functions. We will gradually publish the function library and software on our website http://www.funome.com.

**Discussion**

To find effective markers, features or indicators that can distinguish objective things, here, we developed a funome project. When we use funome to solve practical problems, there are still some points to pay attention to. (1) There are too many math functions in the function library, and researchers need to choose some common, important functions in their study. We will gradually provide some important function sets for downloading and using. (2) For whole genome, transcriptome and proteome data, there were too many markers detected through chips and sequencing. Under the current speed limit of computer, it is impossible to scan all functions for all the markers. We recommend that researchers can select several interesting functions to scan the whole genome or transcriptome.

In the funome study, some pattern recognition methods, such as classification and discrimination, can also be used to improve the efficiency of identifying novel indicator/characteristic.

Although funome comes from the life science, it can be applied to other fields, such as Physics and Economics.

We also hope some excellent indicators/characteristics identified by the funome strategy can be applied in the field of artificial intelligence.

## Acknowledgements

This work was supported in part by the National Natural Science Foundation of China (Grant No. 91746113).